\documentclass[a4paper,11pt]{article}
\usepackage{graphicx} 
\usepackage{float}
\usepackage{amsmath}
\usepackage{pos}
\usepackage{subcaption}
\newcommand{\dd}{\mathrm{d}}
\newcommand{\rt}{\mathbf{r}}
\newcommand{\bt}{\mathbf{b}}
\newcommand{\kt}{\mathbf{k}}
\newcommand{\pt}{\mathbf{p}}
\newcommand{\xbj}{x_\mathrm{Bj}}

\title{Constraining the Dipole Amplitude at NLO with HERA Data and Probing Gluon Saturation in SIDIS}
\ShortTitle{NLO BK Fits and SIDIS}

\author*[a,b]{Carlisle Casuga}
\author*[a,b]{Swaleha Mulani}
\author[a,b]{Heikki M\"antysaari}

\affiliation[a]{Department of Physics, University of Jyv\"askyl\"a,  P.O. Box 35, 40014 University of Jyv\"askyl\"a, Finland}

\affiliation[b]{Helsinki Institute of Physics, P.O. Box 64, 00014 University of Helsinki, Finland}

\emailAdd{carlisle.doc.casuga@jyu.fi}
\emailAdd{swaleha.n.mulani@jyu.fi}
\emailAdd{heikki.mantysaari@jyu.fi}
\date{September 2026}

\abstract{ 
We infer the initial condition for the Balitsky-Kovchegov (BK) evolution equation at next-to-leading order accuracy from HERA proton structure function data. 
A  simultaneous description of 
both  the total cross section and charm  production measurements is obtained. We then show that inclusive hadron production in deep inelastic scattering can provide complementary constraints on the non-perturbative initial condition of the BK equation. As a more differential observable, it serves as a sensitive probe of gluon saturation phenomena. We demonstrate this by calculating predictions for the nuclear modification factor in  hadron production in photon-nucleus scattering at leading order in the dipole picture. We find that saturation effects lead to significant nuclear suppression at small $\xbj$.
}
\begin{document}

\maketitle

\section{Introduction}

Studying gluon saturation is one of the main goals of the Electron Ion Collider (EIC). Highly precise predictions, especially in the small-$x$ region, are needed in anticipation of upcoming measurements. For example, theoretical uncertainties surrounding the non-perturbative input to the high-energy evolution equation must be rigorously estimated and high perturbative order achieved. 

Perturbative evolution equations, such as the Balitsky-Kovchegov (BK) equation~\cite{Kovchegov:1999yj,Balitsky:1995ub}, describe the energy dependence of the scattering processes. In this contribution, we present the Bayesian inference of the initial condition for the BK evolution equation at next-to-leading order accuracy from the proton structure function data, reported in~\cite{Casuga:2026xxt}. For a consistent higher order treatment, the DIS cross sections are computed using both NLO impact factors~\cite{Beuf:2021srj} and the full NLO BK equation~\cite{Balitsky:2007feb}. The resulting posterior distributions represent the uncertainty of the initial condition parameters, and samples from it enable the propagation of this uncertainty to cross sections at small-$x$ computed within the CGC framework. 

Semi-inclusive deep inelastic scattering (SIDIS) provides a more differential probe of the small-$x$ dynamics, with an explicit dependence on the produced hadron transverse momentum $\pt$. Therefore, SIDIS offers a more powerful observable for studying signatures of gluon saturation. The SIDIS cross section at leading order in $\alpha_\mathrm{s}$ in the dipole picture has been computed in \cite{Mueller:1999wm,Kovchegov:2015zha}. In this proceeding, we present predictions for single inclusive hadron production in deep inelastic electron-nucleus scattering. In particular, we demonstrate that saturation effects result in a significant nuclear suppression, quantified in terms of a nuclear modification factor, and that such measurements can also provide complementary constraints for the inference of the BK equation initial condition.


\section{Initial condition for the BK equation at next-to-leading order accuracy}

\begin{figure*}[tb]
    \centering
    \includegraphics[width=0.45\linewidth]{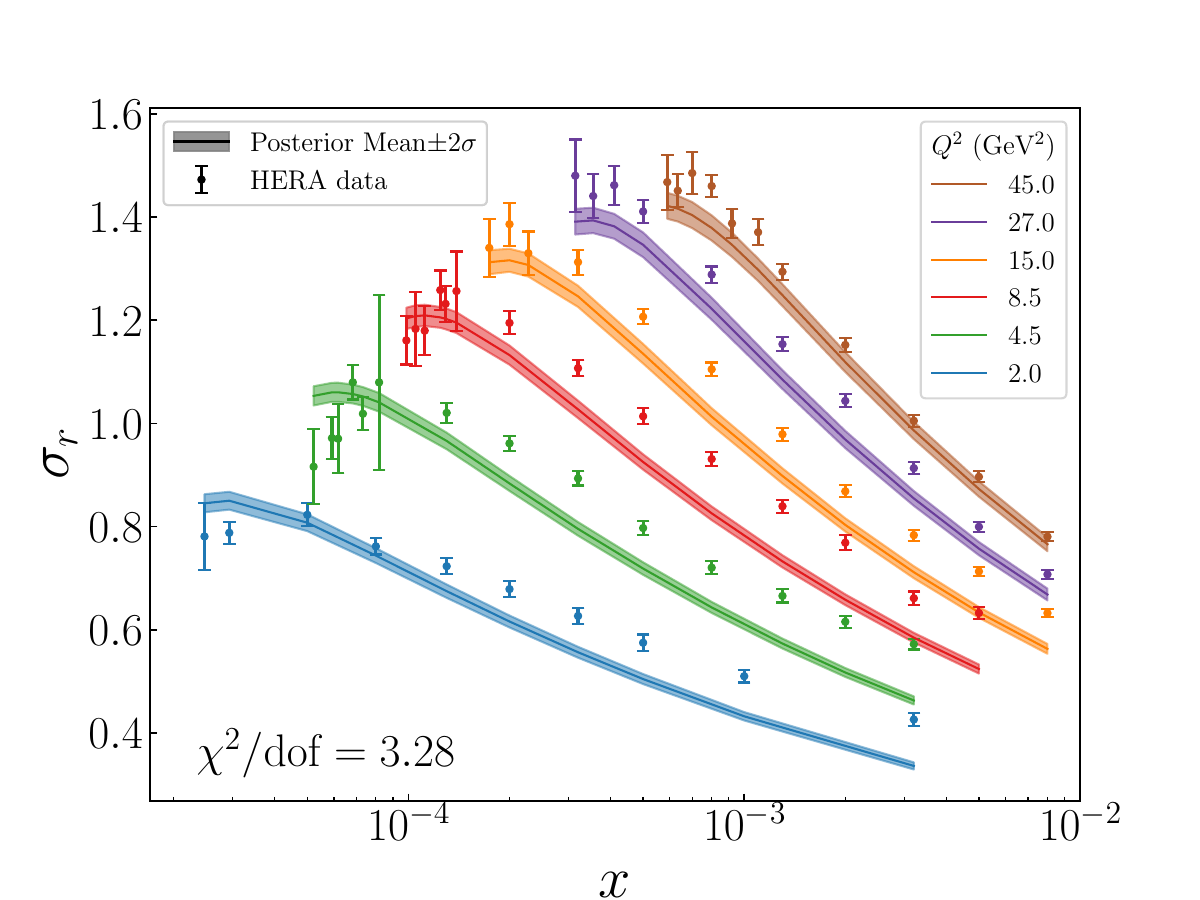}
    \includegraphics[width=0.45\linewidth]{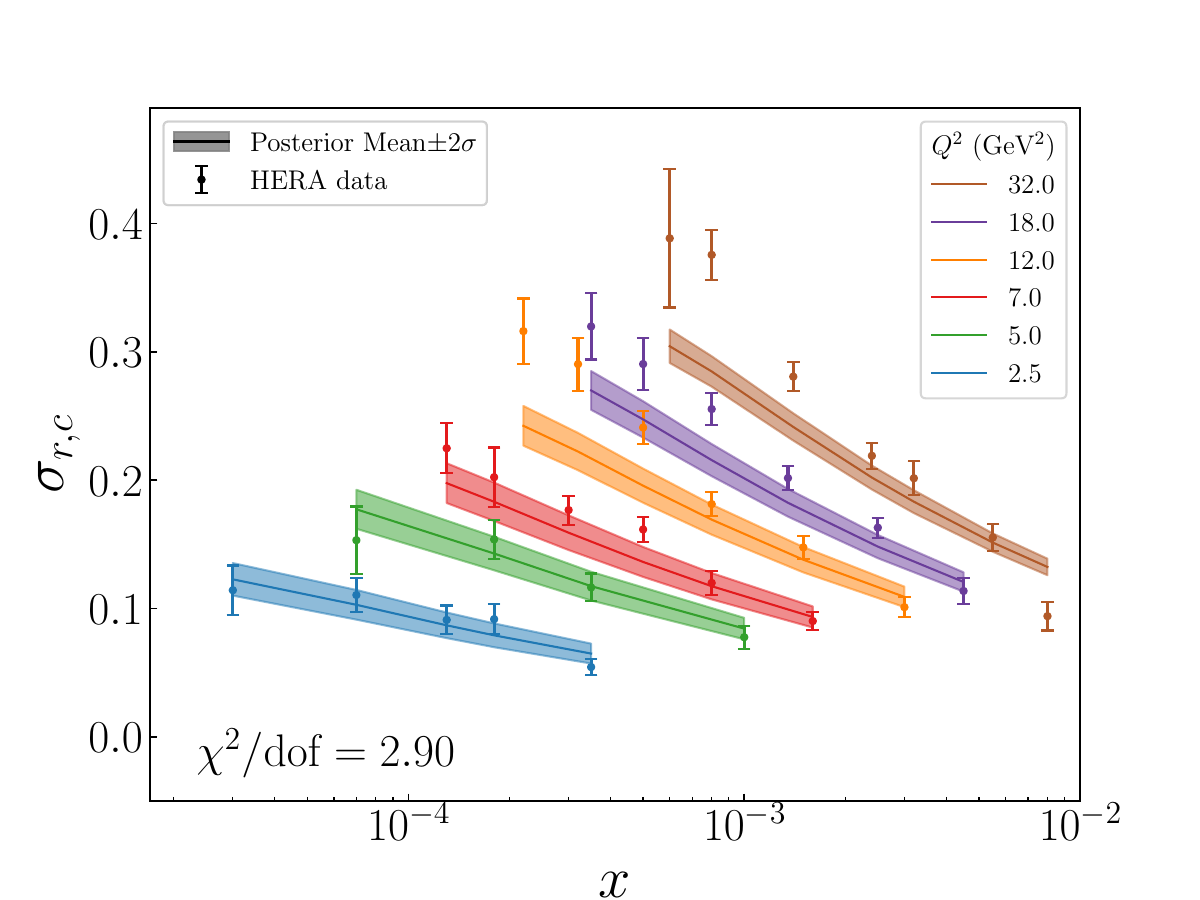}
    \caption{ Comparison between HERA data and total (left) and charm (right) cross section calculations at NLO using the $\mathrm{MV}$ initial condition using Bal+SD running coupling prescription.}
    \label{fig:modelvemulator}
\end{figure*}

The dipole-proton scattering amplitude $N$ at the initial condition is written using an MV-model inspired parametrization
\begin{equation}
        N(\mathbf{r}) = 1-\exp \left[  - \frac{\left(\mathbf{r}^2 Q_{s,0}^2\right)^\gamma}{4} \ln \left( \frac{1}{|\mathbf{r}| \Lambda_\text{QCD}} +  e_c \cdot e   \right)   \right]. 
        \label{eq:initialcondition}
\end{equation}
The free parameter $Q_{s,0}^2$ parametrizes the initial saturation scale, the anomalous dimension $\gamma$ describes the low-$\mathbf{r}$ behavior of the amplitude, and the parameter $e_c$ serves as the infrared regulator. We call the initial condition where we set $\gamma=1$ and $e_c$ to be a free parameter as the ``MV$^e$'' setup. Additionally, the ``MV$^\gamma$'' initial condition would describe the initial condition in Eq. \eqref{eq:initialcondition} where $e_c=1$ and $\gamma$ is a free parameter. The setup where $e_c=\gamma=1$ is the original ``MV'' initial condition.

We extract from the HERA total cross section \cite{H12015} and charm quark contribution \cite{H1:2018flt} measurements the posterior likelihood distribution for the parameters $\boldsymbol{\theta} = \{Q_{s,0}^2, \gamma, C^2, \sigma_0/2, m_c\}$. The HERA data is included  in the small-$x$ region, $x\leq0.01$, within the range 2 GeV$^2 \leq Q^2 \leq$ 45 GeV$^2$. In this fit, we combine NLO DIS hard factors with the full NLO BK equation. Higher order corrections enhanced by large transverse logarithms are also resummed in the NLO BK equation, see~\cite{Casuga:2026xxt} for details.

\begin{figure*}[h!]
    \centering
    \includegraphics[width=0.45\linewidth]{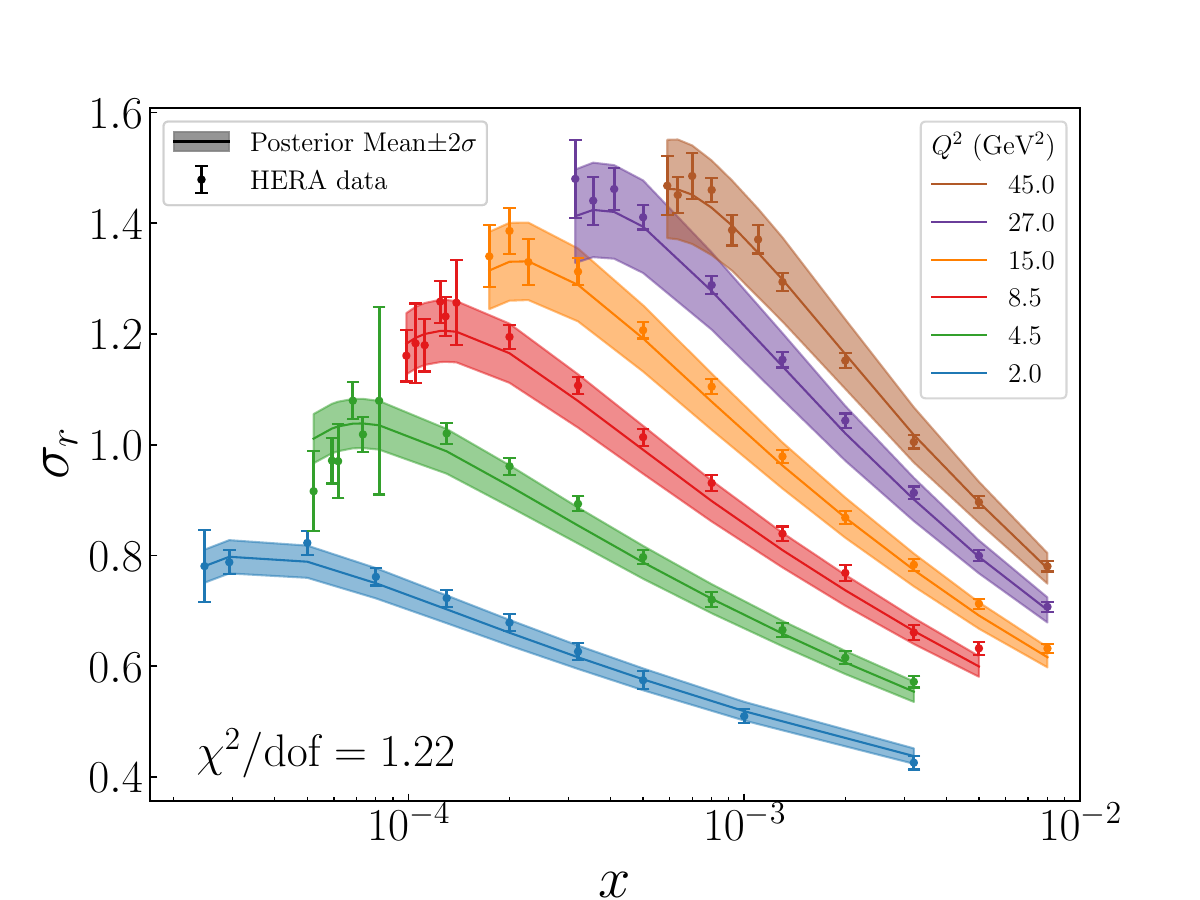}
    \includegraphics[width=0.45\linewidth]{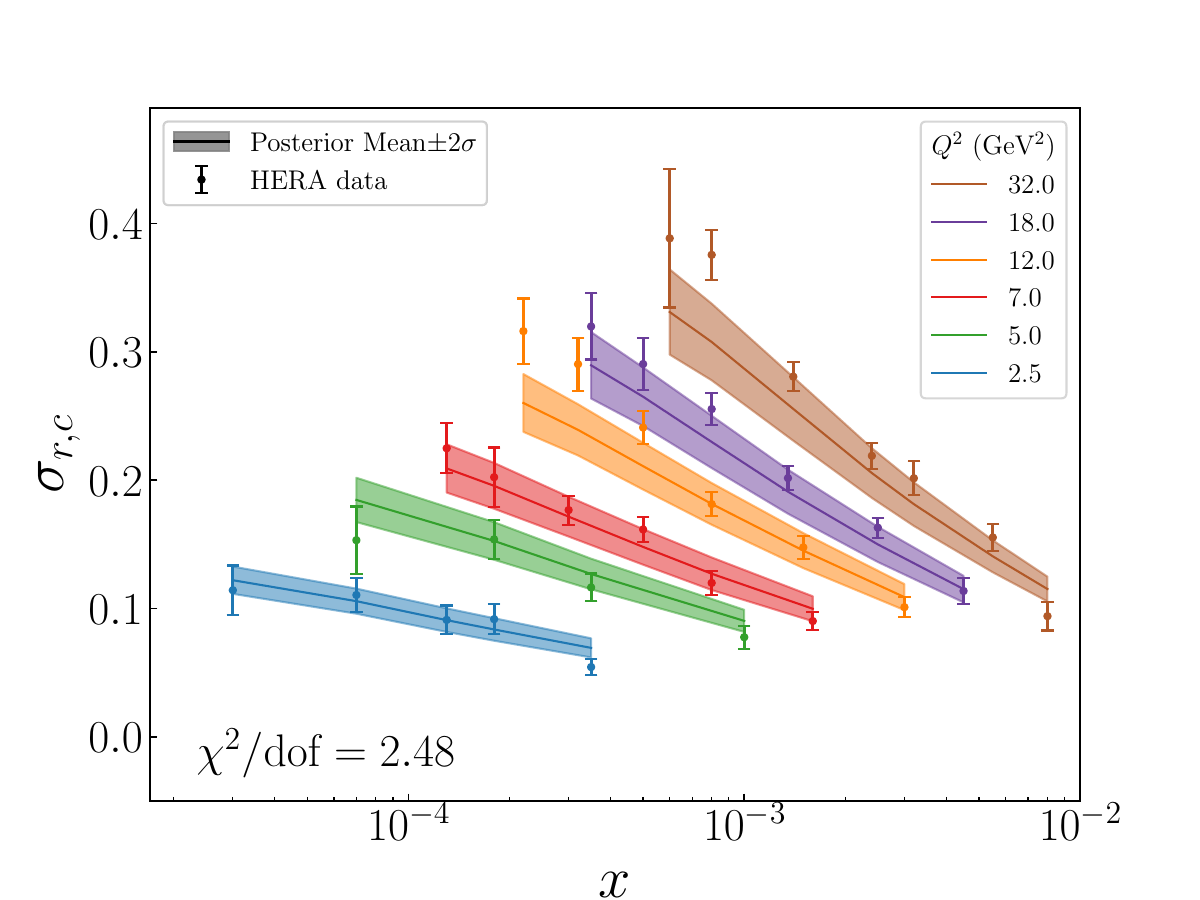}
    \caption{ Same as Fig. \ref{fig:modelvemulator} but with the MV$^\gamma$ initial condition.}
    \label{fig:modelvemulator_mvgam}
\end{figure*}

We first perform the analysis using the MV model initial condition parametrization. This initial condition ensures a positive definite dipole amplitude in the momentum space, however, as evident in Fig.~\ref{fig:modelvemulator}, it does not adequately describe the total cross section data. Next we use the  MV$^\gamma$ model which offers more flexibility, and a successful simultaneous description of both total inclusive cross section and charm quark data is obtained, as seen in Fig.~\ref{fig:modelvemulator_mvgam}.

The median values for the model parameters are presented in Table \ref{tab:medianvalues}. The HERA data constrains the parameters very well. The $C^2$ parametrizes the coordinate space running coupling scale, controlling the evolution speed. HERA data prefers large values of $C^2$ that correspond to slower evolution. The data also favor a steep initial dipole amplitude with $\gamma > 1$. Such a solution, however, does not lead to a positive-definite momentum-space dipole amplitude, which enters directly in observables such as hadron production in SIDIS (see Eq.~\eqref{eq:momspacedipole}).

\renewcommand{\arraystretch}{1.5}
\begin{table}[ht]
    \centering
    \begin{tabular}{|c|c|c|c|c|c|c|}
    \hline
         \textbf{Initial Condition} & $Q_{s,0}^2 \ [\mathrm{GeV}^{2}]$  & $\gamma$ & $C^2$ & $\sigma_0/2 \ $[mb] & $m_c \ [\mathrm{GeV}]$ &  $\chi^2/N_\mathrm{dof}$\\
         \hline
         \hline
         MV & $(7.21^{+0.62}_{-0.60})\cdot 10^{-3}$ 
         & 1 (fixed) & $9.83^{+2.92}_{-2.08}$ & $22.8^{+3.1}_{-2.5}$ & $1.11^{+0.09}_{-0.09}$
         & 3.18 
         \\
         \hline
         MV$^\gamma$ & $0.089^{+0.002}_{-0.005}$ & $4.30^{+0.29}_{-0.11}$ & $915^{+517}_{-361}$ & $22.4^{+1.1}_{-1.1}$ & $1.17^{+0.06}_{-0.05}$& 1.31
         \\          
         \hline
    \end{tabular}
    \caption{Median and $2\sigma$ values of the posterior distributions of the MV and $MV^\gamma$ initial condition fit of NLO DIS to HERA data.}
    \label{tab:medianvalues}
\end{table}

\section{Semi inclusive hadron production in DIS}

The differential cross section for semi inclusive $\gamma^*$-hadron scattering at leading order in the dipole picture is given as,
 \begin{equation}
 \label{eq:hadronic_xsection}
                \frac{\dd\sigma^{\gamma^* h}_{\mathrm{L,T}}}{\dd^2 \pt\, \dd y} \propto  
                \int \dd z_f \;
               \frac{\dd\sigma^{\gamma^* + h \to q + X}_{\mathrm{L,T}}}{\dd(\pt / z_f)\, \dd y}
                \otimes
                \mathcal{D}(\mu^2, z_f).
            \end{equation} 
Here $ \mathcal{D}(\mu^2, z_f)$ is the fragmentation function evaluated at scale $\mu^2=\pt^2$. We  use the leading order fragmentation function NNFF1.0~\cite{Bertone:2017tyb}. It is convoluted with the partonic level cross section, which in momentum space is expressed as,
\begin{align}
\label{eq:partonic_xsection}
    \frac{\dd\sigma^{\gamma^* + h \to q + X}_{\mathrm{L,T}}}{\dd \pt\, \dd y} \propto \sum_f e_f^2 \int \! d^2\bt \int \! d^2\kt  \ {\tilde S(\kt, \bt)} \otimes \mathcal{H}_{\mathrm{L,T}}(\pt, \kt, z_1).
\end{align}
Here, $\mathcal{H}_{\mathrm{L,T}}(\pt, \kt, z_1)$ are the hard factors for longitudinal or transverse polarizations of the photon given in~\cite{Casuga:2026dif}. We compute the differential cross section in momentum space to isolate the numerically challenging momentum space dipole amplitude  $\tilde S$, which is the two-dimensional Fourier transform of the dipole amplitude:
\begin{equation}
\label{eq:momspacedipole}
    \tilde S(\kt,\bt,x_g) =  \int \mathrm{d}^2{\rt} \ e^{i \kt \cdot \rt} \left[ 1-N(\rt,\bt,x_g) \right].
\end{equation}
Here $x_g$ is the longitudinal momentum fraction of the target gluons involved in the scattering and is given in ~\cite{Casuga:2026dif}. The momentum space dipole amplitude $\tilde S$ is evaluated as follows. First, at each BK evolution rapidity we fit the numerical solution to the BK equation using a parametrization 
\begin{equation}
\label{eq:dipolefitparam}
    N_f (\rt,\bt,x_g) = \Bigg\{ 1 - \exp \Biggl[ \Biggl(- \frac{(r^2 Q_{s,0}^2)^\gamma}{4} 
   \ln\left(\frac{1}{r\Lambda_\mathrm{QCD}}+ e_c \cdot e\right) \Biggr)^p \Biggr] \Bigg \}^{\frac{1}{p}}.
\end{equation}
Then, to obtain $ \tilde S(\kt,\bt,x_g)$, the Hankel transform of Eq.~\eqref{eq:dipolefitparam} is computed using standard algorithms~\cite{2019JOSS....4.1397M}.

\begin{figure}[tb]
    \centering
    \includegraphics[width=.45\textwidth]{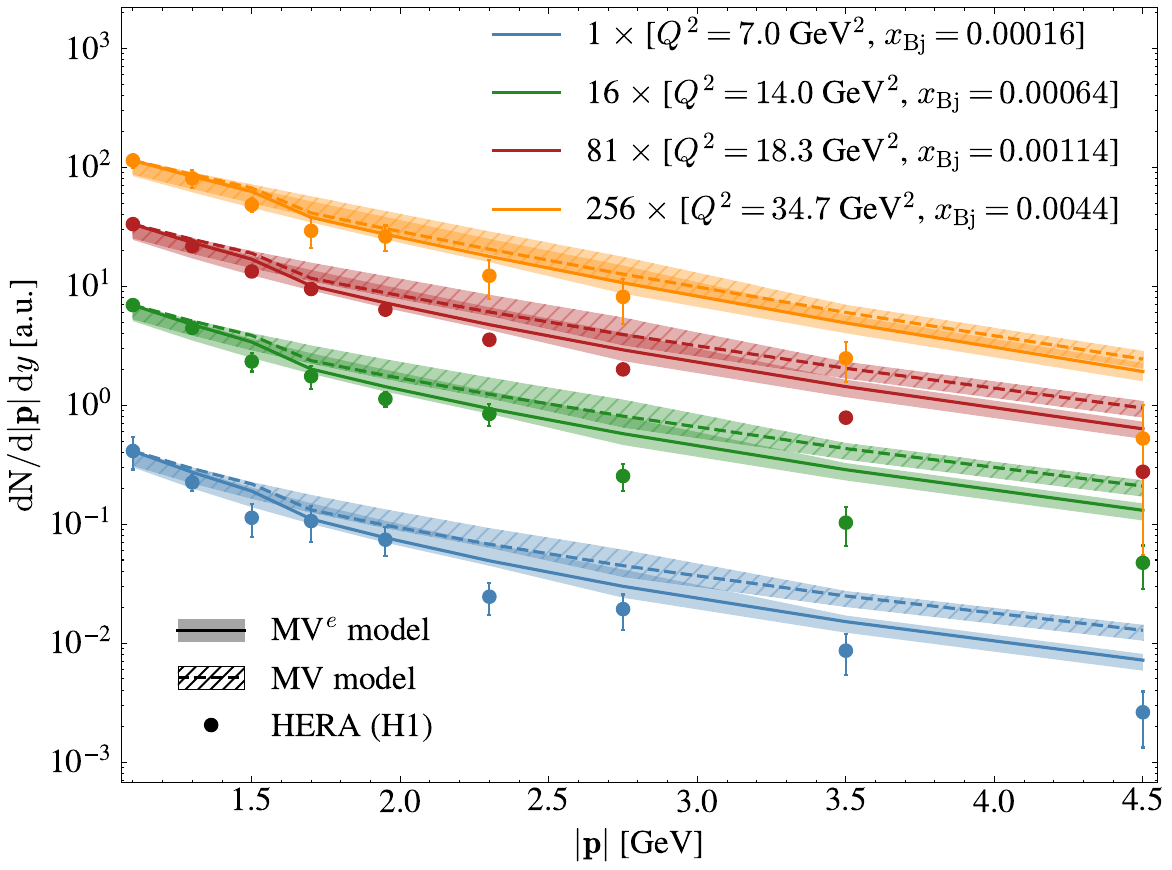}
    \caption{ 
    Charged hadron transverse momentum spectra (arbitrary units) in $\gamma^* + p \to h^\pm + X$ scattering in the photon-proton center-of-mass frame at $y=2$ compared with the HERA data~\cite{H1:1996muf}.
    The uncertainty band corresponds to variation of the fragmentation function scale $\mu$ by factor 2. 
    }
    \label{plot:Plot_FF_Scale_Uncertainty_HERA_3Tables_gP_HERA_VS_MVe_VS_MV}
\end{figure}

We first compare the calculated cross section to the charged hadron transverse momentum spectra from measured by the H1 collaboration~\cite{H1:1996muf}. The results are computed using both the (leading order) MV and MV$^e$ dipole amplitude fits from~\cite{Lappi:2013zma,Casuga:2023dcf}, and are shown in  Fig.~\ref{plot:Plot_FF_Scale_Uncertainty_HERA_3Tables_gP_HERA_VS_MVe_VS_MV}. The hadron spectra are shown with the  fragmentation function  scale uncertainty estimated by varying the scale $\mu$ by a factor of $2$. The spectra is normalized separately at each $Q^2$ value to match with lowest $\pt$ data points of \cite{H1:1996muf}. For lower $\pt$ both the MV and MV$^e$ fits show similar spectra in HERA kinematics and describe the HERA data well. However, at higher $|\pt| \geq 3.5$ GeV, the computed spectra are too hard, especially when MV model dipoles are used. 
%

To study nuclear suppression caused by gluon saturation effects in semi-inclusive hadron production, we compute the nuclear modification factor,
\begin{equation}
\label{eq:RpA}
    R_\mathrm{pA} = \frac{{\dd \sigma^{\gamma^* + Au \to h + X}/(\dd^2\pt \ \dd{z_h}) }}{A \times {\dd \sigma^{\gamma^* + p \to h + X}/(\dd^2\pt \ \dd{z_h})} }.
\end{equation}
In the dilute limit and at the initial condition  of the BK evolution, $ R_\mathrm{pA}$  goes to 1 at high $|\pt|$, which corresponds no nuclear modification in the dilute regime. We obtain $ R_\mathrm{pA} $ by summing over  contributions from longitudinal and transverse polarizations of the incoming virtual photon computed using Eq.~\eqref{eq:hadronic_xsection}. In Eq.~\eqref{eq:RpA}, $\dd \sigma^{\gamma^* + Au \to h + X}/(\dd^2\pt \ \dd{z_h}) $ is the charged pion production cross section for $\gamma^* +$Au scattering and $\dd \sigma^{\gamma^* + p \to h + X}/(\dd^2\pt \ \dd{z_h})$ is the  cross section for $\gamma^* + p$ scattering. For $R_\mathrm{pA} $ predictions, the MV$^e$ fit is used.

\begin{figure*}[ht]
\centering

\centering
\includegraphics[width=0.45\linewidth]{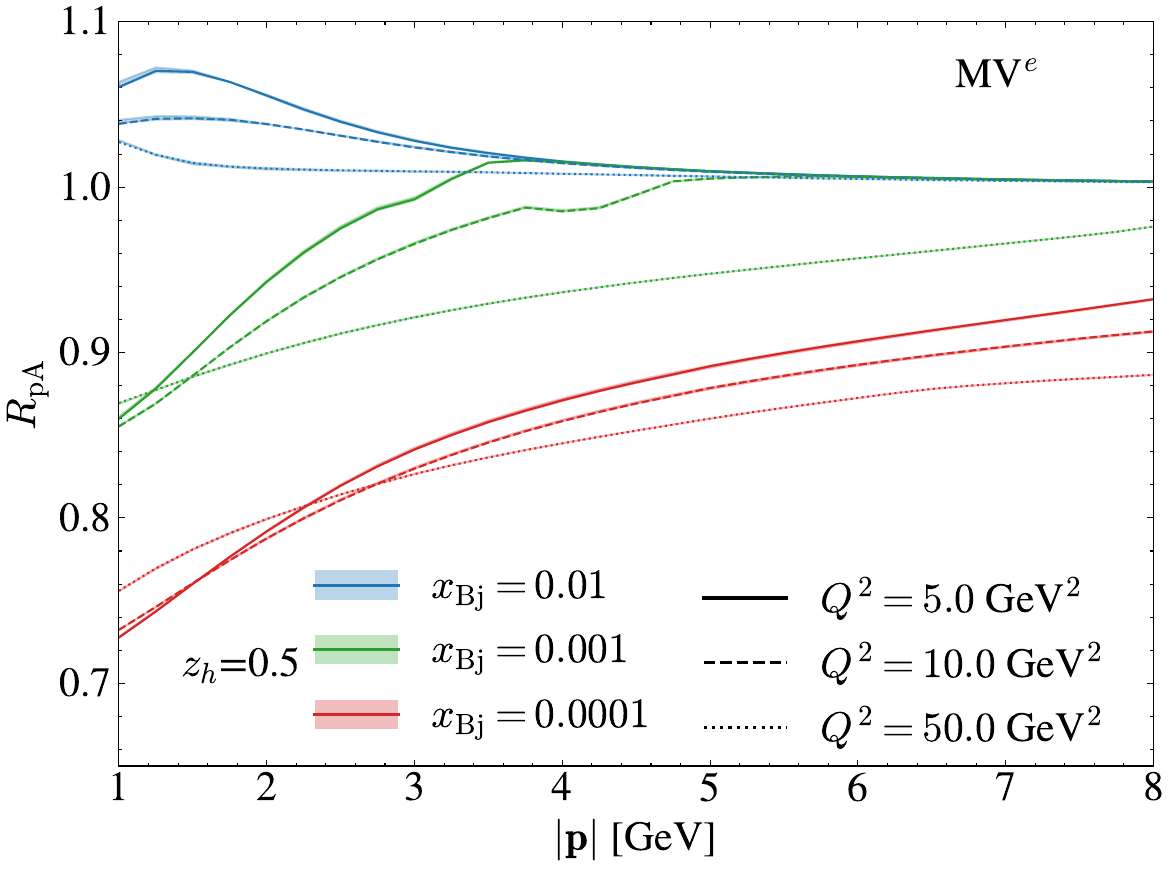}
\includegraphics[width=0.45\linewidth]{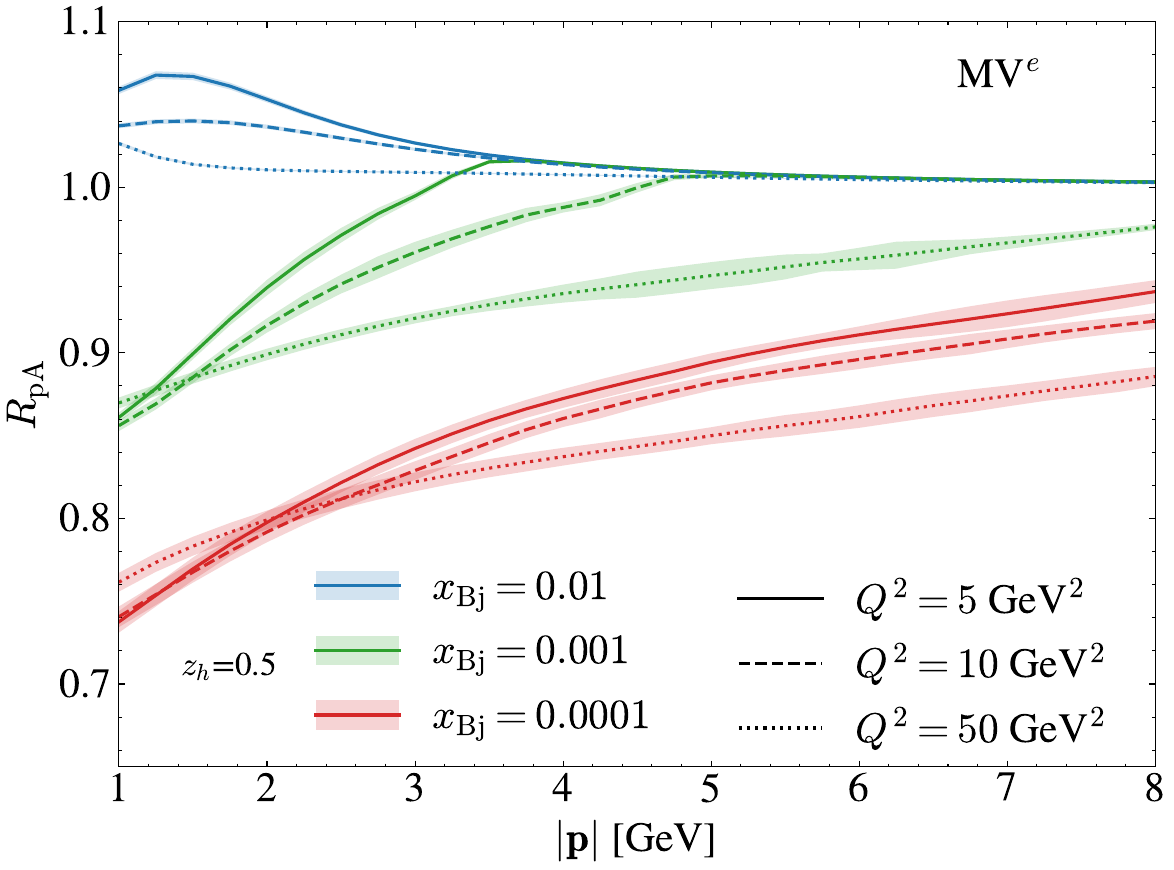}
\caption{Nuclear modification factor as a function of $|\pt|$ for selected values of $Q^2$ and $\xbj$ with fixed $z_h=0.5$. Results are shown for FF scale uncertainty (left) and posterior sample uncertainty in initial conditions for dipole amplitudes (right).
}
\label{fig:Plot_SIDIS_nuclear_mod_EIC_z105_Q2_5_100_mb}
\end{figure*}

In Fig.~\ref{fig:Plot_SIDIS_nuclear_mod_EIC_z105_Q2_5_100_mb}, we show nuclear modification factors $R_\mathrm{pA}$ computed for inclusive $\pi^\pm$ production as function of transverse momenta of pion $\pt$ for selected values of $Q^2$ and $\xbj$. The results are calculated for a fixed fraction of the photon longitudinal momentum carried by produced hadron, $z_h = 0.5$. 
The $R_\mathrm{pA}$ is found to approach  unity for high $\pt$ especially at larger $\xbj$ close to the initial condition of the BK evolution $x_0=0.01$. For lower $\pt$ at $\xbj = 0.01$, we see a pronounced Cronin peak, which is washed out at smaller $\xbj$ as a result of the  BK evolution. Furthermore, with decreasing $\xbj$, more suppression in $R_\mathrm{pA}$ is observed. We found a relatively weak dependence on the photon virtuality $Q^2$. These results are consistent with the results of Ref.~\cite{Lappi:2013zma}, where nuclear modification factor is computed for inclusive hadron production in proton-nucleus scattering in a similar CGC framework.

We estimate two sources of uncertainties for $R_\mathrm{pA}$. The first one arises from the choice of the fragmentation function scale $\mu$, as shown on the left side of  Fig.~\ref{fig:Plot_SIDIS_nuclear_mod_EIC_z105_Q2_5_100_mb}. The second one originates from the parametrization of  initial condition for the BK evolution, shown on the right panel of Fig.~\ref{fig:Plot_SIDIS_nuclear_mod_EIC_z105_Q2_5_100_mb}. The uncertainties due to the scale choice are estimated by varying $\mu$ by a factor 2. In the case of $R_\mathrm{pA}$, this uncertainty is found to be negligible. Uncertainties propagated from the initial condition of the BK evolution are moderate and weakly dependent on $\pt$. We also note that the central values for $R_\mathrm{pA}$, computed by averaging over posterior samples, are almost identical to those obtained using parametrization given in Table I of \cite{Casuga:2026dif} and shown on left panel of Fig.~\ref{fig:Plot_SIDIS_nuclear_mod_EIC_z105_Q2_5_100_mb}.

\section{Summary \& Conclusions}

We have presented best-fit values for the non-perturbative input of the Balitsky-Kovchegov equation, along with uncertainty estimates. By using both the NLO impact factors and the NLO BK equation, a simultaneous fit to and a successful description of both total inclusive and charm quark cross section data is obtained with the $MV^\gamma$-parametrized initial condition. Further studies are required in order to obtain initial conditions that provide a good description of the HERA data and have a positive definite Fourier transform required e.g. when calculating single inclusive cross sections.

As an application, we have computed single inclusive hadron production in deep inelastic scattering at leading order, because as a  more differential observable it can be expected to be a more sensitive probe of the gluon saturation phenomena.
We validate our results against HERA SIDIS data, and a good agreement is obtained  especially at small $\pt$.   We have  computed predictions for the nuclear modification factor that can be measured at the EIC. Our results indicate significant nuclear suppression due to gluon saturation in inclusive pion production, which is more pronounced towards smaller $\xbj$. The uncertainty arising due to choice of the fragmentation function scale almost vanishes at the nuclear modification factor level, while uncertainties propagated from the initial condition parametrization of dipole amplitude are still visible. This demonstrates that future EIC measurements might provide complementary constraints on the initial condition of the BK initial condition inferred from experimental data.

\acknowledgments
This work was supported by the Research Council of Finland, the Centre of Excellence in Quark Matter, and projects 338263 and 359902, and by the European Research Council (ERC, grant agreements  No. ERC-2023-101123801 GlueSatLight and ERC-2018-ADG-835105 YoctoLHC).
The support of the Vilho, Yrjö and Kalle Väisälä Foundation is also acknowledged.
Computing resources from CSC – IT Center for Science in Finland and the Finnish Grid and Cloud Infrastructure (persistent identifier \texttt{urn:nbn:fi:research-infras-2016072533}) were used in this work.
The content of this article does not reflect the official opinion of the European Union and responsibility for the information and views expressed therein lies entirely with the authors.

\bibliographystyle{JHEP}
\bibliography{refs.bib}

\end{document}